\documentclass[10pt,a4paper]{article}
\usepackage[utf8]{inputenc}
\usepackage{amsmath}
\usepackage{amsfonts}
\usepackage{amssymb}
\usepackage{graphicx}
\usepackage{lmodern}
\usepackage{subcaption}
\usepackage{authblk}
\usepackage{cite}
\usepackage{url}

\title{Kerr-Newman black hole surrounded by quintessence under quantum gravity effects and gravity's rainbow.}
\author{
   Aheibam Boycha Meitei$ ^{1\dagger} $, Irom Ablu Meitei$ ^{1\ddagger} $,  Telem Inungochouba Singh$ ^{2\dagger \dagger} $, and Kangujam Yugindro Singh$ ^{1\ast} $\\
 $ ^{1} $ {\small Department of Physics, Manipur University,\\ Canchipur,
Imphal,\\ Manipur 795003, India.}\\
$ ^{2} $ {\small Department of Mathematics, Manipur University,\\ Canchipur,
Imphal,\\ Manipur 795003, India.}\\
$ ^{\dagger} ${\small \texttt{aheibamboycha143@gmail.com}}\\
$ ^{\ddagger} ${\small \texttt{ablu.irom@gmail.com}}\\
$ ^{\dagger \dagger} ${\small \texttt{ibungochouba@rediffmail.com}}\\

$ ^{\ast} ${\small \texttt{yugindro361@gmail.com}}
\date{}
}

\begin{document}
\maketitle

\begin{abstract}
In this paper, we investigate the quantum gravity effects on the tunneling of particles and gravity's rainbow across the event horizon of Kerr-Newman black hole (KNBH) surrounded by the quintessence. The aspect of quantum gravity has been introduced by applying the Generalized Uncertainty Principle (GUP) to the Klein-Gordon equation and the Dirac equation of scalar and fermion particles. By solving the Generalized Klein-Gordon and Dirac equations obeyed by scalar and fermion particles, corrections to the Hawking temperature of the KNBH in the presence of quintessence is discussed. The tunneling of fermions is also examined using the modified Hamilton-Jacobi equation also known as modified Rarita-Schwinger equation, and the corrected Hawking temperature is determined. The corrected Hawking temperature of the KNBH surrounded by quintessence is found to be dependent not only on the properties of the black hole but also on the quantum numbers of the emitted particles and quintessence. Then, we explored the KNBH surrounded by quintessence within the framework of gravity rainbow using rainbow functions. In the context of rainbow functions in loop quantum gravity, we derive the Hawking temperature, heat capacity, equation of state parameters, and entropy for the KNBH surrounded by quintessence. It is found that these quantities are influenced by both the quintessence and the rainbow gravity.
\end{abstract}



\section{Introduction}	
Hawking demonstrated that black holes are not entirely “black”; instead, they emit radiation similar to that of a black body, with a temperature that relates to the surface gravity of the black hol\cite{1,2} . Such radiation is due to quantum effects, and it is called Hawking radiation. Hawking radiation has now become an important aspect of the study of black holes. Hawking's black hole area theorem and the discovery of Hawking radiation using quantum field theory in curved space-time, show that there might be some relationship between the properties of the black hole and thermodynamics\cite{3,4,5} . Bekenstein\cite{6,7} and Bardeen\cite{8} also pointed out several similarities between black hole physics and thermodynamics. Various methods have been proposed to study Hawking radiation as the tunneling of particles.

In the radial null geodesic method proposed by Kraus et al.\cite{9,10} a potential barrier is created by the outgoing particles. The Semi-classical WKB method proposed by Parikh-Wilczek\cite{11} is used to obtain the imaginary part of the action. It is observed that when the back reaction of the emitted particles is taken into account, the emitted spectrum deviates from the precise thermal spectrum.
Angheben et al.\cite{12} used the relativistic Hamilton-Jacobi equation and the WKB approximation to study Hawking radiation as tunneling for the extremal and rotating black holes without considering the back reaction of the emitted particles. This method, which is also known as the Hamolton-Jacobi method, is considered as an extension of the complex path analysis developed by Padmanabhan et al.\cite{13,14,15} .

 Kerner and Mann\cite{16} introduced the study of tunneling of Dirac particles for general stationary black holes by choosing appropriate Gamma matrices. Using the Dirac equation and the semi-classical WKB approximation, the action of the emitted Dirac particle, which is related to the Boltzmann factor for emission at Hawking temperature, is obtained. The tunneling of fermions for various black holes is studied in Refs.~\cite{17,18,19,20,21,22}. Refs.~\cite{23,24,25,26} used the tortoise coordinate transformation in the study of Hawking radiation. Using the WKB approximation to the Proca equation, Refs.~\cite{27,28,29} investigated the tunneling of vector boson particles across the event horizon of black hole.

Various theories of quantum gravity e. g., string theory, loop quantum gravity, doubly special theory of relativity, and Gedanken experiments\cite{30,31,32,33,34,35,36,37} predicted the existence of a minimum length. This minimal length may be considered as the minimum position uncertainty\cite{38} , and the Generalized Uncertainty Principle (GUP) can be obtained through the modified commutation relation. The Generalized Klein-Gordon equation, Generalized Dirac Equation, and Generalized Proca equation are obtained by using the GUP to the Klein-Gordon equation, Dirac Equation, and Proca equation respectively. The effects of GUP on the tunneling of scalar and fermion particles are discussed in Refs.\cite{39,40,41,42,43,44,45,46,47} .

Lorentz symmetry is a symmetry that requires the laws of physics to be the same for all observers\cite{48} . It provides the dispersion relation of Einstein's theory of special relativity, which is considered a fundamental relation in both general relativity and quantum field theory. However, quantum theories of gravity, such as string theory\cite{49} and loop quantum gravity\cite{50} , predict the violation of Lorentz symmetry at the Planck energy scale, $ E_{p}\simeq 10^{19}GeV $. Lorentz symmetry violation can be given in the form of modified dispersion relations (MDRs). The theory discussed about the modified dispersion relations is known as double special relativity (DSR), and it is considered an extension of the special theory of relativity\cite{51,52,53} . In DSR, the rainbow gravity(RG) is the generalization of the rainbow spacetime, where the spacetime metric depends on the energy of the probing particle to solve the nonlinearity issue of the Lorentz transformation\cite{54}. Research has been conducted on the impact of the gravity rainbow within the context of the Friedmann-Robertson-Walker universe\cite{55,56,57,58,59}, black hole thermodynamics\cite{60,61,62,63,64,65,66,67,68,69,70} , and the remnants of black holes\cite{71,72,73}. Cosmological eras such as inflation and scale-invariant fluctuations are also studied in the context of gravity rainbow\cite{74,75}. Several other papers also studied compact objects under different gravity models. Considering three different cases for bulk viscous fluids, Yousaf\cite{76} investigated its coefficients in the background of four dimensional Einstein–Gauss–Bonnet gravity. Ref.\cite{77} solved the equations for radial perturbations of neutron stars in the four dimensional Einstein–Gauss–Bonnet gravity theory with Skyrme Lyon or Brussels-Montreal Skyrme functionals class equations of state, along with the Müller-Serot equations of state. The Joule–Thomson expansion and thermodynamic characteristics of charged anti-de Sitter black holes are examined in the context of Cotton gravity coupled to nonlinear electrodynamics in the paper\cite{78}.

Recent astronomical observations indicate that the expansion of the universe is accelerating. It is believed to be driven by what is known as the dark energy\cite{79,80,81} . The origin of the dark energy is still elusive. Different dark energy models are proposed e.g. the cosmological constant, quintessence energy etc. The static Cosmological constant is characterized by the equation of state $\omega=-1$, while quintessence is dynamic with an equation of state $\omega=\frac{p}{\rho}$ where $p$ is the pressure and $\rho$ is the energy density in the range $-1<\omega<-\frac{1}{3}$. Kiselev\cite{82} introduced an exact solutions of the Einstein equations that are static and spherically symmetric, incorporating quintessential matter around either a charged black hole or one that is uncharged, as well as in scenarios without a black hole. Refs.\cite{83,84,85,86} developed static, spherically symmetric solutions for black holes or droplets by defining a fuzzy dark matter density profile, such as the Einasto density profile or Zhao parameterization. In this paper, we study the effects of quantum gravity on the tunneling of scalar and fermion particles, as well as the impact of gravity's rainbow on the thermodynamics of a KNBH surrounded by quintessence. The entropy correction of the black hole is also discussed for the tunneling of fermion beyond the semi classical approximation. The combination of GUP and Rainbow Gravity is theoretically valid, as they originate at different structural levels within the quantum gravity theory. GUP arises from modified canonical commutation relations, leading to a minimal measurable length and quantum corrections to phase space and particle dynamics, whereas, Rainbow Gravity is based on modified dispersion relations that introduce energy dependence into the spacetime geometry which a test particle experiences. Both GUP and Doubly Special Relativity (which led to Rainbow gravity) suggest modification of commutation relation between position and momenta \cite{86a,86b}. Analogous to the minimum measurable length in GUP, Doubly Special Relativity is associated with a maximum energy scale leading to breaking of Lorentz invariance eventually \cite{86c}.  In principle, a complete theory of quantum gravity, if exists, should be able to integrate the two quantum gravity-inspired models. The presence of both GUP and Rainbow gravity, in a single manuscript, will illustrate the broader approach of quantum gravity phenomenology. However, currently, there is no fundamental theory of quantum gravity that unifies the phase-space deformation induced by the GUP with the energy-dependent geometry of rainbow gravity. So, these two quantum gravity-inspired models can be treated independently and this approach may be helpful in integrating them. In this paper, we analyse GUP and rainbow gravity separately to investigate the impact of each quantum gravity model on the thermodynamic properties of the Kerr-Newman black hole surrounded by quintessence. 

The paper is organised as follows: In section 2, the KNBH in the presence of quintessence is studied. In section 3, we investigate the quantum tunneling of KNBH in the presence of quintessence using Generalized Klein-Gordon equation and calculate the Hawking temperature of the black hole. In section 4, we investigate the quantum tunneling of fermions from KNBH surrounded by quintessence using the Generalized Dirac Equation. In section 5, using the modified Hamilton-Jacobi equation, we calculate the modified Hawking temperature of the black hole surrounded by quintessence. The correction of Hawking temperature and entropy of the black hole beyond the semi-classical theory is also calculated under the modified Hamilton-Jacobi equation.  In section 6, we investigate the impact of gravity's rainbow on the thermodynamics of the KNBH surrounded by quintessence.  Discussion and conclusion are given in section 7.  

\section{KNBH in the presence of quintessence}
In the Boyer-Lindquist coordinates $(t,r,\theta,\phi)$, the KNBH metric in the Kiselev quintessence is\cite{87,88}
\begin{eqnarray}
ds^{2}=&-&\left( 1-\frac{2Mr-Q^{2}+\alpha r^{1-3\omega}}{\Sigma^{2}}\right) dt^{2}+\frac{\Sigma^{2}}{\Delta}dr^{2}\cr &-&\frac{2a \sin^{2}\theta (2Mr-Q^{2}+\alpha r^{1-3\omega})}{\Sigma^{2}}d\phi dt+\Sigma^{2}d\theta^{2}\cr &+& \sin^{2}\theta \left( r^{2}+a^{2}+a^{2}\sin^{2}\theta \frac{2Mr-Q^{2}+\alpha r^{1-3\omega}}{\Sigma^{2}}\right) d\phi^{2},
\end{eqnarray}
where 
\begin{eqnarray}
\Delta &=&r^{2}-2Mr+a^{2}+Q^{2}-\alpha r^{1-3\omega},\cr
\Sigma^{2}&=&r^{2}+a^{2}\cos^{2}\theta.
\end{eqnarray}
M, a, $\alpha$ and Q represent the black hole mass, angular momentum per unit mass, quintessence parameter  and electric charge of the black hole respectively. The quintessence field parameter $ \omega $ which is defined by the equation of state $ \omega=\frac{p}{\rho} $  $(-1<\omega<-\frac{1}{3})  $, where $ p $ and $ \rho $ are the pressure and density. The smaller value of the quintessence field parameter $ \omega $ gives the greater accelerating effect.

The components of the electromagnetic vector potential are given by
\begin{eqnarray}
A_{t}=-\frac{Q r}{\Sigma^{2}},\ A_{r}=0,\ A_{\theta}=0,\ A_{\phi}=\frac{Q r a \sin^{2}\theta}{\Sigma^{2}}.
\end{eqnarray}

For our convenience in studying of the tunneling process of the black hole black at the event horizon, we perform dragging coordinate transformation\cite{89,90} as 
\begin{eqnarray}
\Omega&=&\frac{d\phi}{dt}=-\frac{g_{03}}{g_{33}},\cr
&=&\frac{a(2Mr-Q^{2}+\alpha r^{1-3\omega})}{(r^{2}+a^{2})\Sigma^{2}+a^{2}\sin^{2}\theta(2Mr-Q^{2}+\alpha r^{1-3\omega})}.
\end{eqnarray} 
Then, Eq. (1) becomes
\begin{eqnarray}
ds^{2}=-\frac{\Delta}{r^{2}+a^{2}+a^{2}\sin^{2}\theta(\frac{2Mr-Q^{2}+\alpha r^{1-3\omega}}{\Sigma^{2}})}dt^{2}+\frac{\Sigma^{2}}{\Delta}dr^{2}+\Sigma^{2}d\theta^{2} .
\end{eqnarray}
With this frame dragging, it is convenient for the study of the tunneling process at the event horizon, as the event horizon of the black hole coincides with the infinite red-shift surface. Hence, the geometrical optical limit becomes a trustworthy estimation, and the WKB approximation can be applied.

The non-vanishing component of vector potential corresponding to Eq. (5) is
\begin{eqnarray}
A_{0}=-\frac{Qr(r^{2}+a^{2})}{(r^{2}+a^{2})\Sigma^{2}+a^{2}\sin^{2}\theta(2Mr-Q^{2}+\alpha r^{1-3\omega})}.
\end{eqnarray}

Considering the emission of radiation along the axis of rotation $ \theta=0,\pi $, the angular velocity and the electromagnetic potential near the event horizon of the black hole become
\begin{eqnarray}
\Omega=\frac{a}{(r^{2}_{h}+a^{2})}
\end{eqnarray}
and
\begin{equation}
A_{0}=-\frac{Qr_{h}}{(r_{h}^{2}+a^{2})}.
\end{equation}
If we consider the upper limiting value of the quintessence field parameter $ \omega=-\frac{1}{3} $, the event horizon of the KNBH is given by 
\begin{eqnarray}
r_{\pm}=\frac{M\pm\sqrt{M^{2}-(1-\alpha)(a^{2}+Q^{2})}}{1-\alpha},
\end{eqnarray}
where $ r_{+} $ and $ r_{-} $ symbolize the outer and inner horizon of the black hole respectively.

The surface gravity of the KNBH near the event horizon in the presence of quintessence is given by\cite{91}
\begin{eqnarray}
\kappa &=& \lim_{g_{00}\to 0} \left( -\frac{1}{2}\sqrt{-\frac{g^{11}}{g_{00}}}\frac{dg_{00}}{dr}\right),\cr &=& \frac{r_{h}-M-\frac{r_{h}^{-3\omega}\alpha (1-3\omega)}{2}}{r_{h}^{2}+a^{2}},
\end{eqnarray}

Using the relation $ T_{H}=\frac{\kappa}{2\pi} $, the Hawking temperature of the KNBH in the presence of quintessence can be obtained as
\begin{eqnarray}
T_{H}= \frac{r_{h}-M-\frac{r_{h}^{-3\omega}\alpha (1-3\omega)}{2}}{2\pi(r_{h}^{2}+a^{2})}.
\end{eqnarray} 
Here, $ M $ is the mass parameter of the KNBH in the presence of quintessence, and it is given by
\begin{eqnarray}
M=\frac{1}{2}\left( r_{h}+\frac{a^{2}}{r_{h}}+\frac{Q^{2}}{r_{h}}-\alpha r_{h}^{-3\omega}\right) .
\end{eqnarray}
\section{Quantum tunneling of scalar particles from KNBH in the presence of quintessence}
The existence of a minimal length of the order of Planck length is a feature of various approaches of quantum gravity, such as string theory, loop quantum gravity, doubly special theory of relativity, and Gedanken experiments. This minimal length leads to a modification of Heisenberg uncertainty principle to a Generalized Uncertainty Principle (GUP) through a modified commutation relation $ [x_{\mu},p_{\nu}]=i\hbar\delta_{\mu\nu}[1+\beta p^{2}] $, where $ x_{\mu} $ and $p_{\nu}  $ are the position and momentum operators. The expression for the GUP can be written as\cite{92}
\begin{eqnarray}
\Delta x\Delta p\geq\frac{\hbar}{2}\left[ 1+\beta(\Delta p)^{2}\right],
\end{eqnarray}
where $\beta=\beta_{0}\frac{l_{p}^{2}}{\hbar^{2}}$, $ \beta_{0} $ is a dimensionless parameter of order unity and $l_{p}=\sqrt{\frac{G\hbar}{c^{3}}}$ is the Planck length.

Incorporating the Generalized Uncertainty Principle (GUP), we can write the generalized Klein-Gordon equation in the presence of quintessence as\cite{41}
\begin{eqnarray}
&-&(-i\hbar\partial^{t}-qA^{0})(-i\hbar\partial_{t}-qA_{0})\Psi=\Big[ (-i\hbar\partial^{\mu}-qA^{\mu})(-i\hbar\partial_{\mu}-qA_{\mu})+m^{2}\Big]\cr &\times& \Big[1-2\beta\lbrace (-i\hbar\partial^{\mu}-qA^{\mu})(-i\hbar\partial_{\mu}-qA_{\mu}) +m^{2}\rbrace\Big]\Psi.
\end{eqnarray}
We consider the tunneling of scalar particles from the event horizon of the KNBH surrounded by the quintessence. To decouple the radial and angular parts from Eq. (14), the wave function of the scalar particles can be written as 
\begin{equation}
\Psi=exp\left[ \frac{i}{\hbar}I(t,r,\theta)\right],
\end{equation}
where $ I(t,r,\theta) $ is the action of the scalar particles.

Using the wave function and the inverse matrices from the line element given in Eq.(5) into the generalized Klein-Gordon equation, we obtain the generalized Hamilton-Jacobi equation 
\begin{eqnarray}
-g^{00}\Big(\frac{\partial I}{\partial t}-qA_{0}\Big)^{2}=\Big[g^{11}\Big(\frac{\partial I}{\partial r}\Big)^{2}+g^{22}\Big(\frac{\partial I}{\partial \theta}\Big)^{2}+m^{2}\Big]\cr\times\Big[1-2\beta\lbrace g^{11}\Big(\frac{\partial I}{\partial r}\Big)^{2}+g^{22}\Big(\frac{\partial I}{\partial \theta}\Big)^{2}+m^{2}\rbrace \Big].
\end{eqnarray}
It is difficult to get the action of the tunneling particles directly from the above equation. Considering the properties of the space-time of the black hole, we carry out the separation of variables on the action $ I $ as
\begin{equation}
I=-(\varepsilon-j\Omega)t+R(r)+W(\theta)+K,
\end{equation}
where $ \varepsilon $ is the energy of the emitted particle, $ j $ the angular momentum and $ K $ a constant parameter.

Then Eq. (16) reduces to
\begin{eqnarray}
A(\partial_{r}R)^{4}+B(\partial_{r}R)^{2}+C=0
\end{eqnarray}
where
\begin{eqnarray}
A&=&-2\beta(g^{11})^{2},\cr
B&=&g^{11}-4\beta g^{11}g^{22}\Big(\frac{\partial W}{\partial \theta}\Big)^{2}-4m^{2}\beta g^{11},\cr
C&=&-2\beta (g^{22})^{2}\Big(\frac{\partial W}{\partial \theta}\Big)^{4}+g^{22}\Big(\frac{\partial W}{\partial \theta}\Big)^{2}-4\beta m^{2}g^{22}\Big(\frac{\partial W}{\partial \theta}\Big)^{2}\cr&& -2\beta m^{4}+m^{2}+g^{00}(-\varepsilon +j\Omega-qA_{0})^{2}.
\end{eqnarray}
Solving Eq. (18) for R(r), we obtain
\begin{eqnarray}
R_{\pm}&=&\pm\int \frac{(r^{2}+a^{2})}{\Delta}(\varepsilon-j\Omega+qA_{0})\cr&&\Big[1+\frac{\beta (r^{2}+a^{2})(\varepsilon-j\Omega+qA_{0})^{2}}{\Delta}\Big]dr.
\end{eqnarray}
Here $ R_{+} $ and $ R_{-} $ represent the outgoing and ingoing solutions. Near the event horizon of the black hole we use the approximation $ \Delta(r)\approx (r-r_{h})\Delta_{r}^{'}(r_{h}) $. Applying residue theorem at the pole $ r=r_{h} $ and considering the emission of radiation along the axis of rotation $ \theta =0,\pi $, we get
\begin{eqnarray}
R_{\pm}=\pm \pi i \frac{(r_{h}^{2}+a^{2})(\varepsilon-j\Omega_{h}+qA_{0})}{\left( r_{h}-M-\frac{\alpha (1-3\omega)r_{h}^{-3\omega}}{2}\right) }\Big[1+\Pi\beta\Big],
\end{eqnarray}
where $\Pi= \frac{r_{h}^{-2-3\omega}\Big[3r_{h}\alpha \omega\left\lbrace 3a^{2}\omega +r_{h}^{2}(2+3\omega) \right\rbrace -r_{h}^{3\omega}\left\lbrace a^{4}+3Q^{2}r^{2}_{h}+a^{2}(Q^{2}+4r^{2}_{h}) \right\rbrace  \Big](\varepsilon-j\Omega+qA_{0})^{2}}{2\Big(r_{h}-M-\frac{\alpha (1-3\omega)r_{h}^{-3\omega}}{2}\Big)^{2}} $.

Using the WKB approximation the tunneling probabilities of the particles across the event horizon of the black hole are
\begin{equation}
\Gamma_{emission}=exp[-2(ImR_{+}+ImK)]
\end{equation}
and
\begin{equation}
\Gamma_{absorption}=exp[-2(ImR_{-}+ImK)].
\end{equation}
The tunneling rate of scalar particles across the event horizon of the black hole is
\begin{eqnarray}
\Gamma_{rate}&=&exp[-2(ImR_{+}-ImR_{-})]\cr
&=&exp\Big[\frac{-2\pi (r_{h}^{2}+a^{2})(-\varepsilon+j\Omega_{h}-qA_{0})}{(r_{h}-M-\frac{\alpha (1-3\omega)r_{h}^{-3\omega}}{2})}[1+\Pi\beta]\Big].
\end{eqnarray}
The tunneling rate is related to the Boltzmann factor for emission at the Hawking temperature\cite{11}. The Hawking temperature is obtained as
\begin{eqnarray}
T_{BH}&=&\frac{(r_{h}-M-\frac{\alpha (1-3\omega)r_{h}^{-3\omega}}{2})}{2\pi(r_{h}^{2}+a^{2})}[1-\Pi\beta]\cr &=&T_{H}[1-\Pi\beta].
\end{eqnarray}
where $ T_{H} $ is the original Hawking temperature of the KNBH in the presence of quintessence without any quantum gravity correction. The corrected Hawking temperature $ T_{BH} $ increases or decreases according to $ \Pi <0 $ or $ \Pi >0 $ . 

\section{Quantum tunneling of fermions from KNBH in the presence of quintessence}
We employ the generalized Dirac equation to examine the Hawking temperature for fermions by taking into account the tunneling process in the context of quantum gravity effects. The Generalized Dirac equation is given by \cite{42,93,94}

\begin{eqnarray}
\left(i\gamma^{j} \partial_{j} -\frac{e}{\hbar}\gamma^{\mu}A_{\mu}+\frac{m}{\hbar}\right)\left(1-\beta m^{2}+\beta\hbar^{2}\partial_{k}\partial^{k} \right)\Psi = -i\gamma^{0}\partial_{0}\Psi,   
\end{eqnarray} 
where $ j,k=1,2,3 $ and $ \mu=0,1,2,3 $ represent the spatial coordinates.

We can express Eq. (5) as
\begin{eqnarray}
ds^{2}=-g_{tt}dt^{2}+g_{rr}dr^{2}+g_{\theta\theta}d\theta^{2}.
\end{eqnarray}
From Eq. (27), the contravariant components are given by
\begin{eqnarray}
g^{tt}&=&\frac{r^{2}+a^{2}+a^{2}\sin^{2}\theta(\frac{2Mr-Q^{2}+\alpha r^{1-3\omega}}{\Sigma^{2}})}{\Delta},\cr
g^{rr}&=&\frac{\Delta}{\Sigma^{2}},\ \ g^{\theta\theta}=\frac{1}{\Sigma^{2}}.
\end{eqnarray}
Then, the gamma matrices are chosen as
\begin{eqnarray}
\gamma^{t}&=&\sqrt{g^{tt}}\begin{pmatrix}
i & 0\\
0 & -i
\end{pmatrix},\ \ \gamma^{r}=\sqrt{g^{rr}}\begin{pmatrix}
0 & \sigma^{3}\\
\sigma^{3} & 0
\end{pmatrix},\cr
\gamma^{\theta}&=&\sqrt{g^{\theta\theta}}\begin{pmatrix}
0 & \sigma^{1}\\
\sigma^{1} & 0
\end{pmatrix},
\end{eqnarray}
where $ \sigma^{i} $ $(i=1,2,3)$ are the Pauli matrices. 

For a spin -1/2 fermion, two states exist that correspond to spin up and spin down. In this work, we focus on the spin-up case. The scenario for spin down is analogous to that of spin up. To describe the motion of fermions through the black hole's event horizon, the ansatz wave function for the fermion in the spin-up state is given by
\begin{eqnarray}
\Psi=\begin{pmatrix}
X\\
0\\
Y\\
0
\end{pmatrix}exp\left( \frac{i}{\hbar}I(t,r,\theta,\phi)\right), 
\end{eqnarray}
where $ I $ is the action of the emitted fermion and $ X $ and $ Y $ are the functions of $ (t,r,\theta,\phi) $.

Incorporating the wave function and gamma matrices into the generalized Dirac equation, applying the WKB approximation, and retaining only the first order of $ \hbar $, we derive the four decoupled equations   
\begin{eqnarray}
-i\sqrt{g^{tt}}(\partial_{t}I)X+\Big[1-\beta m^{2}-\beta\left\lbrace g^{rr}(\partial_{r}I)^{2}+g^{\theta\theta}(\partial_{\theta}I)^{2} \right\rbrace \Big]\cr \times \Big[(m-i e A_{0}\sqrt{g^{tt}})X-\sqrt{g^{rr}}(\partial_{r}I)Y \Big]=0,
\end{eqnarray}
\begin{eqnarray}
\Big[1-\beta m^{2}-\beta\left\lbrace g^{rr}(\partial_{r}I)^{2}+g^{\theta\theta}(\partial_{\theta}I)^{2} \right\rbrace \Big]\left\lbrace -\sqrt{g^{\theta\theta}}(\partial_{\theta}I)Y \right\rbrace =0,
\end{eqnarray}
\begin{eqnarray}
i\sqrt{g^{tt}}(\partial_{t}I)Y+\Big[1-\beta m^{2}-\beta\left\lbrace g^{rr}(\partial_{r}I)^{2}+g^{\theta\theta}(\partial_{\theta}I)^{2} \right\rbrace \Big]\cr \times \Big[(m+i e A_{0}\sqrt{g^{tt}})Y-\sqrt{g^{rr}}(\partial_{r}I)X \Big]=0,
\end{eqnarray}
\begin{eqnarray}
\Big[1-\beta m^{2}-\beta\left\lbrace g^{rr}(\partial_{r}I)^{2}+g^{\theta\theta}(\partial_{\theta}I)^{2} \right\rbrace \Big]\left\lbrace -\sqrt{g^{\theta\theta}}(\partial_{\theta}I)X \right\rbrace =0.
\end{eqnarray}
From Eq. (41) and Eq. (43), we get
\begin{eqnarray}
\Big[1-\beta m^{2}-\beta\left\lbrace g^{rr}(\partial_{r}I)^{2}+g^{\theta\theta}(\partial_{\theta}I)^{2} \right\rbrace \Big]\left\lbrace -\sqrt{g^{\theta\theta}}(\partial_{\theta}I) \right\rbrace =0.
\end{eqnarray}
The correction parameter $ \beta $ is a very small quantity. Therefore, Eq.(35) reduces to 
\begin{eqnarray}
 -\sqrt{g^{\theta\theta}}(\partial_{\theta}I)  =0 \cr
 \Rightarrow g^{\theta\theta}(\partial_{\theta}I)^{2}=0
\end{eqnarray}
Considering the properties of the black hole, we carry out the separation of variable on the action of the fermion as 
\begin{eqnarray}
I=-(E-j \Omega)t+W(r,\theta)+j \phi +K, 
\end{eqnarray}
where $ E $ is the energy of the emitted fermion, $ j $ the angular momentum and $ K $ a constant parameter.

Substituting Eq. (36) and Eq. (37) into Eq. (31) and Eq. (33), we obtain
\begin{eqnarray}
\Big[i(E-j\Omega)\sqrt{g^{tt}}+\left\lbrace 1-\beta m^{2}-\beta g^{rr}(\partial_{r}W)^{2} \right\rbrace \left( m-i e A_{0}\sqrt{g^{tt}}\right)   \Big]X \cr -\sqrt{g^{rr}}(\partial_{r}W)\left\lbrace 1-\beta m^{2}-\beta g^{rr}(\partial_{r}W)^{2} \right\rbrace Y=0
\end{eqnarray}
\begin{eqnarray}
-\sqrt{g^{rr}}(\partial_{r}W)\left\lbrace 1-\beta m^{2}-\beta g^{rr}(\partial_{r}W)^{2} \right\rbrace X+\Big[i(E-j\Omega)\sqrt{g^{tt}}\cr +\left\lbrace 1-\beta m^{2}-\beta g^{rr}(\partial_{r}W)^{2} \right\rbrace \left( m+i e A_{0}\sqrt{g^{tt}}\right)   \Big]Y=0
\end{eqnarray}
The non-trival solution of the two Eq. (38) and Eq. (39) can be derived when the determinant of the two coefficients of $ X $ and $ Y $ is zero. Ignoring higher power of $ \beta $, we derive the radial equation of motion for the emitted fermion particle as 
\begin{eqnarray}
\left\lbrace g^{tt}e^{2}A^{2}_{0}+m^{2}-g^{rr}(\partial_{r}W)^{2} \right\rbrace \Big[ 1-2\beta\left\lbrace m^{2}+g^{rr}(\partial_{r}W)^{2} \right\rbrace   \Big]\cr -2 g^{tt}e A_{0}(E-j\Omega)\Big[ 1-\beta\left\lbrace m^{2}+g^{rr}(\partial_{r}W)^{2} \right\rbrace \Big]+g^{tt}(E-j\Omega)^{2}=0.
\end{eqnarray}
Eq. (40) can be put in the form of a biquadratic equation as
\begin{eqnarray}
a(\partial_{r}W)^{4}+b(\partial_{r}W)^{2}+c=0,
\end{eqnarray}
where
\begin{eqnarray}
a&=&2\beta(g^{rr})^{2},\cr
b&=&g^{rr}\left\lbrace 2\beta g^{tt}eA_{0}(E-j\Omega-eA_{0})-1)\right\rbrace, \cr
c&=&g^{tt}(E-j\Omega-eA_{0})^{2}+m^{2}-2\beta m^{2}\left\lbrace m^{2}-g^{tt}eA_{0}(E-j\Omega-eA_{0})\right\rbrace .
\end{eqnarray}
Solving Eq. (41) for W, we obtain
\begin{eqnarray}
W_{\pm}&=&\pm\int\frac{(r^{2}+a^{2})(E-j\Omega -e A_{0})}{\Delta}\cr&& \Big[1+\beta \frac{(r^{2}+a^{2})(E-j\Omega)(E-j\Omega-eA_{0})}{\Delta} \Big]dr.
\end{eqnarray}
$ W_{+} $ and $ W_{-} $ represent the outgoing and ingoing fermion particles. In the vicinity of the black hole's event horizon, we employ the approximation $ \Delta(r)\approx (r-r_{h})\Delta_{r}^{'}(r_{h}) $. By applying residue theorem and Feynman prescription at the pole $ r=r_{h} $ and examining the radiation emission along the rotational axis $ \theta =0,\pi $, we arrive 
\begin{eqnarray}
W_{\pm}=\pm \pi i \frac{(r_{h}^{2}+a^{2})(E-j\Omega_{h}-eA_{0})}{\left( r_{h}-M-\frac{\alpha (1-3\omega)r_{h}^{-3\omega}}{2}\right) }\Big[1+\Upsilon\beta\Big],
\end{eqnarray}
where $\Upsilon= \frac{r_{h}^{-2-3\omega}\Big[3r_{h}\alpha \omega\left\lbrace 3a^{2}\omega +r_{h}^{2}(2+3\omega) \right\rbrace -r_{h}^{3\omega}\left\lbrace a^{4}+3Q^{2}r^{2}_{h}+a^{2}(Q^{2}+4r^{2}_{h}) \right\rbrace  \Big](E-j\Omega-eA_{0})(E-j\Omega)}{2\Big(r_{h}-M-\frac{\alpha (1-3\omega)r_{h}^{-3\omega}}{2}\Big)^{2}} $.

According to WKB approximation the tunneling rate of the particles across the event horizon of the black hole is
\begin{eqnarray}
\Gamma_{rate}&=&exp[-2(ImW_{+}-ImW_{-})]\cr
&=&exp\Big[\frac{-2\pi (r_{h}^{2}+a^{2})(E-j\Omega_{h}-eA_{0})}{\left( r_{h}-M-\frac{\alpha (1-3\omega)r_{h}^{-3\omega}}{2}\right) }[1+\Upsilon\beta]\Big].
\end{eqnarray}
This tunneling rate is related to the Boltzmann factor for emission at the Hawking temperature\cite{11}. The modified Hawking temperature is obtained as
\begin{eqnarray}
T_{QF} &=&\frac{\left( r_{h}-M-\frac{\alpha (1-3\omega)r_{h}^{-3\omega}}{2}\right) }{ 2\pi(r_{h}^{2}+a^{2})}\Big[1-\Upsilon\beta \Big]\cr
&=&T_{H}\Big[1-\Upsilon\beta \Big],
\end{eqnarray}
where $ T_{H} $ is the Hawking temperature of the Kerr-Newman black hole surrounded by the quintessence without quantum correction as specified in Eq. (11). $T_{QF}$ increases or decreases according to $ \Upsilon <0 $ or $\Upsilon >0  $.
\section{Modified Hamilton Jacobi Equation of Fermion Tunneling}
The deformed Lorentz dispersion relation in the study of string theory and quantum gravity at the Planck energy scale is given by [51,52,95-100] 
\begin{eqnarray}
E^{2}=\vec{P}^{2}+m^{2}-(\lambda E)^{n}\vec{P}^{2},
\end{eqnarray}
where $ E $ and $\vec{P}$ are the energy and momentum of the particle of static mass $ m $ and $ \lambda $ is a constant on the Planck energy scale. To derive the modified Dirac equation from Eq. (47), the value of $ n $ is taken as unity in the Liouville string theory. Ref.~\cite{99} also studied the modified Dirac equation for $ n=2 $.

By using the deformed Lorentz dispersion relation, the modified Rarita-Schwinger equation is derived within a gravitational context, and subsequently, the modified Hamilton-Jacobi equation is obtained through a semiclassical approximation\cite{101,102} . 
\begin{eqnarray}
g^{\mu\nu}(\partial_{\mu}S+e A_{\mu})(\partial_{\nu}S+e A_{\nu})+m^{2}-2\sigma m g^{00}(\epsilon-eA_{t})^{2}=0,
\end{eqnarray}
where $ \epsilon $ is the energy of the radiating particles and $ S $ is the action of the radiation particles defined by 
\begin{eqnarray}
S=-\epsilon t+R_{0}(r)+\Re(\theta)+j\phi,
\end{eqnarray}
where $ \partial_{r}S=-\omega $ and $ \partial_{\phi}S=j $. As the modification is on the quantum scale, we assume the coupling constant as $ \sigma\ll 1 $.

From Eq. (1) contravariant metric components are
\begin{eqnarray}
g^{tt}&=&-\frac{1}{\Delta \Sigma^{2}}\left[\left(r^{2}+a^{2} \right)^{2}-a^{2}\Delta \sin^{2}\theta  \right], \cr
g^{t \phi}&=&-\frac{a\left(2M-Q^{2}+\alpha r^{1-3\omega} \right) }{\Sigma^{2}\Delta},\cr
g^{rr}&=&\frac{\Delta}{\Sigma^{2}},\cr
g^{\theta\theta}&=&\frac{1}{\Sigma^{2}},\cr
g^{\phi\phi}&=&\frac{\Delta-a^{2}\sin^{2}\theta}{\Delta \Sigma^{2}\sin^{2}\theta}.
\end{eqnarray}

Using the contravariant metric components of Eq. (50) and Eq. (49) in Eq. (48), we get the modified Hamilton Jacobi equation of the Kerr-Newman black hole surrounded by the quintessence as
\begin{eqnarray}
&&-\frac{1}{\Delta \Sigma^{2}}\left[\left(r^{2}+a^{2} \right)^{2}-a^{2}\Delta \sin^{2}\theta  \right](\epsilon -e A_{t})^{2}+\frac{\Delta}{\Sigma^{2}}(\partial_{r}R_{0})^{2}+\frac{1}{\Sigma^{2}}(\partial_{\theta}\Re)^{2}\cr&& +\frac{\Delta-a^{2}\sin^{2}\theta}{\Delta \Sigma^{2}\sin^{2}\theta}(j+eA_{\phi})^{2}+\frac{2a\left(2M-Q^{2}+\alpha r^{1-3\omega} \right) }{\Sigma^{2}\Delta}(\epsilon -e A_{t})(j+eA_{\phi}) \cr&& +m^{2}+\frac{2\sigma m}{\Delta \Sigma^{2}}\left[\left(r^{2}+a^{2} \right)^{2}-a^{2}\Delta \sin^{2}\theta  \right](\epsilon -e A_{t})^{2}=0
\end{eqnarray}
The radiation emission along the axis at angles $ \theta=0,\pi $ is taken into account, we derive the radial equation motion for the emitted particle as
\begin{eqnarray}
\partial_{r}R_{0}^{\pm}=\pm\frac{(r^{2}+a^{2})}{\Delta}(\epsilon-\epsilon_{0})\sqrt{1-2\sigma m\Theta^{2}},
\end{eqnarray}
where, $ \epsilon_{0}=eA_{t}+\frac{aj}{r^{2}+a^{2}} $ and $ \Theta =\frac{\epsilon-e A_{t}}{\epsilon-\epsilon_{0}} $.

Integrating Eq. (52), the action of the emitted particle can be express as
\begin{eqnarray}
R_{0}^{\pm}=\pm \int\frac{(r^{2}+a^{2})}{\Delta}(\epsilon-\epsilon_{0})\sqrt{1-2\sigma m\Theta^{2}}\ dr,
\end{eqnarray}
where $ R_{0}^{+} $ and $ R_{0}^{-} $ represents the outgoing and ingoing particle.

Near the event horizon $ r=r_{h} $, we used the approximation $\Delta(r)\approx (r-r_{h})\Delta_{r}^{'}(r_{h})  $. Applying residue theorem at the pole $ r=r_{h} $ in Eq.(53), we get
\begin{eqnarray}
R_{0}^{\pm}=\pm\pi i\frac{(r^{2}_{h}+a^{2})(\epsilon-\epsilon_{0})}{\left( r_{h}-M-\frac{\alpha (1-3\omega)r_{h}^{-3\omega}}{2}\right)}(1-\sigma m\Theta^{2}).
\end{eqnarray}
The tunneling rate of the emitted fermion across the event horizon of the black hole is 
\begin{eqnarray}
\Gamma_{rate}&=&exp[-2(ImR_{0}^{+}-ImR_{0}^{-})]\cr
&=&exp\Big[\frac{-2\pi (r_{h}^{2}+a^{2})(\epsilon- \epsilon_{0})}{\left( r_{h}-M-\frac{\alpha (1-3\omega)r_{h}^{-3\omega}}{2}\right) }(1-\sigma m\Theta^{2})\Big].
\end{eqnarray}
This tunneling rate is related to the Boltzman factor for emission at Hawking temperature. The modified Hawking temperature is given by 
\begin{eqnarray}
T_{RH}&=&\frac{\left( r_{h}-M-\frac{\alpha (1-3\omega)r_{h}^{-3\omega}}{2}\right)}{2\pi (r_{h}^{2}+a^{2})}(1+\sigma m\Theta^{2})\cr
&=&T_{H}(1+\sigma m\Theta^{2}).
\end{eqnarray}
For $ \sigma>0 $, the modified Hawking temperature of the KNBH surrounded by the quintessence, as derived from this deformed Rarita-Schwinger Hamilton Jacobi equation, is determined to be greater than the original Hawking temperature given in Eq. (11). 

To achieve more precise outcomes and demonstrate the impact of quantum corrections, we implement the following expansion correction on particle energy and radial action. The particle energy corrected to\cite{103,104,105} 
\begin{eqnarray}
E=E_{0}+\sum_i \hbar^{i} E_{i},
\end{eqnarray}
and the corrected radial action is
\begin{eqnarray}
R=R_{0}+\sum_i \hbar^{i} R_{i}(r),
\end{eqnarray}
where, $ i=1,2,3... $ and $ E_{0}=\epsilon -\epsilon_{0} $. $ E_{0} $ and $ R_{0} $ are the semi classical part of $ E $ and $ R $.

Rewriting and solving Eq. (51) and Eq. (52), we can get the following equations when $ i $ takes different values
\begin{eqnarray}
\hbar^{0};\frac{d R_{0}^{\pm}}{dr}\mid_{r\rightarrow r_{h}}=\pm\frac{E_{0}}{d},
\end{eqnarray}
\begin{eqnarray}
&&\hbar^{1};\frac{d R_{1}^{\pm}}{dr}\mid_{r\rightarrow r_{h}}=\pm\frac{E_{1}}{d},...
\end{eqnarray}
where
\begin{eqnarray}
d=\frac{\left( r_{h}-M-\frac{\alpha (1-3\omega)r_{h}^{-3\omega}}{2}\right)}{(r_{h}^{2}+a^{2})}\left(1+\sigma m \Theta^{2} \right). 
\end{eqnarray}

If we consider $\xi_{i}$ to be the proportional coefficient connecting $ R_{0} $ and $ R_{i} $, we can express Eq. (58) as
\begin{eqnarray}
R^{\pm}=R_{0}^{\pm}+\sum_{i}\xi_{i}\hbar^{i}R_{0}^{\pm}=R_{0}^{\pm}\left(1+\sum_{i}\zeta_{i}\frac{\hbar^{i}(r_{h}^{2}+a^{2})^{i}}{(r_{h}^{2}+a^{2})^{i+1}} \right), 
\end{eqnarray}
where $ \zeta_{i}=\xi_{i}(r_{h}^{2}+a^{2}) $. Substituting $ R_{0}^{\pm} $ into Eq. (59), we get a more accurate tunneling rate as
\begin{eqnarray}
\Gamma_{rate}^{'}&=&exp\left[-2\left(Im R^{+}-Im R^{-} \right)  \right) \cr
&=& exp\Big[\frac{-2\pi(r_{h}^{2}+a^{2})(\epsilon-\epsilon_{0})}{\left( r_{h}-M-\frac{\alpha (1-3\omega)r_{h}^{-3\omega}}{2}\right)} (1-\sigma m \Theta^{2})\cr&& \times\left(1+ \sum_{i}\zeta_{i}\frac{\hbar^{i}(r_{h}^{2}+a^{2})^{i}}{(r_{h}^{2}+a^{2})^{i+1}} \right) \Big] \cr &=& exp\left( -\frac{\epsilon-\epsilon_{0}}{T_{RH}^{'}}\right), 
\end{eqnarray}
where 
\begin{eqnarray}
T_{RH}^{'}=\frac{\left( r_{h}-M-\frac{\alpha (1-3\omega)r_{h}^{-3\omega}}{2}\right)}{2\pi(r_{h}^{2}+a^{2})}(1+\sigma m \Theta^{2})\left(1- \sum_{i}\zeta_{i}\frac{\hbar^{i}(r_{h}^{2}+a^{2})^{i}}{(r_{h}^{2}+a^{2})^{i+1}} \right).
\end{eqnarray}
Eq. (64) represents the more accurate corrected Hawking temperature beyond the semi-classical theory of the black hole surrounded by the quintessence. It is closely related to the correction of black hole entropy.

The Bekenstein-Hawking entropy $ S_{bh} $ of the Kerr-Newman black hole is given by
\begin{eqnarray}
S_{bh}=\pi(r_{h}^{2}+a^{2}). 
\end{eqnarray}
Thermal flux of particles from the black hole due to quantum processes implies the thermodynamic behaviour of the black hole. To calculate the change in entropy, we consider a quasi-static process during which a stationary black hole of mass $ M $, angular momentum $ J $, surface area $ A $ and angular velocity $ \Omega $ is taken to a new stationary black hole with parameters $ M+\delta M $, $ J+\delta J $, $ A+\delta A $ and $ \Omega+\delta \Omega $. According to the first law of black hole thermodynamics the changes in mass, angular momentum, area and the angular velocity are given by
\begin{eqnarray}
dM=T dS +\Omega dJ+\Phi dQ,
\end{eqnarray}
where $ S $ and $ T $ are the entropy and the temperature of the black hole. The angular velocity $ \Omega$ and the electromagnetic potential $ \Phi$ are constant over the event horizon of the black hole.

Using Eq. (64) in Eq. (66), we can get the corrected entropy as
\begin{eqnarray}
S_{BH}&=&\int d S_{BH}=\int \frac{dM-\Omega dJ -\Phi dQ}{T_{RH}^{'}}\cr &=&\int \frac{dM-\Omega dJ -\Phi dQ}{T_{H}}(1-\sigma m \Theta^{2})\left(1+ \sum_{i}\zeta_{i}\frac{\hbar^{i}(r_{h}^{2}+a^{2})^{i}}{(r_{h}^{2}+a^{2})^{i+1}} \right)\cr &=&\int d S_{bh}(1-\sigma m \Theta^{2})\left(1+ \sum_{i}\zeta_{i}\frac{\hbar^{i}(r_{h}^{2}+a^{2})^{i}}{(r_{h}^{2}+a^{2})^{i+1}} \right).
\end{eqnarray} 
From Eq. (59) and Eq. (60)
\begin{eqnarray}
\hbar^{0};\int dS_{bh}(1-\sigma m \Theta^{2})=S_{bh}(1-\sigma m \Theta^{2}),
\end{eqnarray}
\begin{eqnarray}
\hbar^{1}; \int dS_{bh}(1-\sigma m \Theta^{2})\frac{\zeta_{1}}{(r_{h}^{2}+a^{2})}&=&\pi \zeta_{1}(1-\sigma m \Theta ^{2})\ln S_{bh}\cr &=&\zeta_{1}^{'}(1-\sigma m \Theta ^{2})\ln S_{bh},
\end{eqnarray}
Continuing in the same manner, we get
\begin{eqnarray}
\hbar^{2};\int dS_{bh}(1-\sigma m \Theta^{2})\frac{\zeta_{2}}{(r_{h}^{2}+a^{2})}&=&\pi \zeta_{1}(1-\sigma m \Theta ^{2})\ln S_{bh}\cr &=&\zeta_{2}^{'}(1-\sigma m \Theta ^{2})\ln S_{bh}
\end{eqnarray}
\[
\begin{aligned}
    &\vdots \qquad ; \qquad \ldots
\end{aligned}
\]

where $ \zeta_{i}^{'}=\pi \zeta_{i} $. Substituting Eq.(68), Eq.(69), and Eq.(70) into Eq.(67), we get the corresponding more accurate black hole entropy as
\begin{eqnarray}
S_{BH}=\left(S_{bh}+\zeta_{1}^{'}\ln S_{bh}+\zeta_{2}^{'}\ln S_{bh}+.\ .\ . \right)(1-\sigma m \Theta ^{2}). 
\end{eqnarray}
Eq. (71) represents the more accurate entropy of the black hole beyond semi-classical theory.

\section{KNBH surrounded by quintessence in gravity's rainbow}
Since DSR relies on a nonlinear Lorentz transformation in momentum space, the definition of dual position space experiences a nonlinearity from the Lorentz transformation\cite{71,72} . Magueijo and Smolin\cite{106} discussed this problem by proposing that the spacetime background experienced by a test particle depends upon its energy. Consequently, there exists a one-parameter family of metrics that depends on the energy of these test particles, referred to as rainbow metrics.

The nonlinearity of the Lorentz transformation results in a modified dispersion relation as
\begin{eqnarray}
E^{2}f(E/E_{p})^{2}-p^{2}g(E/E_{p})^{2}=m^{2},
\end{eqnarray}
where $ E_{p} $ is the Planck energy scale, $ m $ is the mass of the test particle, $ f(E/E_{p}) $ and $ g(E/E_{p}) $ are rainbow functions, and they satisfy the condition
\begin{eqnarray}
\lim_{E\to 0}f(E/E_{p})= \lim_{E\to 0}g(E/E_{p})=1.
\end{eqnarray}
The RG corrected metric of the KNBH surrounded by quintessence in frame dragging coordinate is given by
\begin{eqnarray}
ds^{2}=\frac{g_{00}}{f(E/E_{p})^{2}}dt^{2}+\frac{g_{11}}{g(E/E_{p})^{2}}dr^{2}+\frac{g_{22}}{g(E/E_{p})^{2}}d\theta^{2}+\frac{g_{33}}{g(E/E_{p})^{2}}d\phi^{2},
\end{eqnarray}
where $ g_{00},g_{11} $, $ g_{22} $ and $ g_{33} $ are the covariant component of the KNBH metric.

The modified Hawking temperature in rainbow gravity is given by\cite{62,63,64}
\begin{eqnarray}
T_{RG}=\frac{g(E/E_{p})}{f(E/E_{p})}T_{H}.
\end{eqnarray}
As stated in references\cite{70,71,72,107,108,109}, the lower bound on the energy of a particle released by a black hole through Hawking radiation is given by 
\begin{eqnarray}
E\geq \frac{1}{\Delta x}\sim \frac{1}{r_{h}}.
\end{eqnarray}
Using Eq. (11) and Eq. (76) in Eq. (75), we obtain
\begin{eqnarray}
T_{RG}=\frac{r_{h}-M-\frac{r_{h}^{-3\omega}\alpha (1-3\omega)}{2}}{2\pi(r_{h}^{2}+a^{2})}\frac{g(1/r_{h}E_{p})}{f(1/r_{h}E_{p})}.
\end{eqnarray}

At energies much larger than the particle mass but smaller than the Planck energy scale, one of the interesting class of modified dispersion relation is of the type\cite{110,111}
\begin{eqnarray}
m^{2}\approx E^{2}-p^{2}+\eta p^{2}\left(\frac{E}{E_{p}} \right)^{n} ,
\end{eqnarray}
with real $ \eta $ of order 1 and integer $n (> 0)$. This equation aligns with several outcomes derived from the Loop-Quantum-Gravity framework and also some findings from theories in $\kappa$-Minkowski noncommutative space time.
\begin{figure}[htbp]
    \centering
    \subfloat[]{%
        \includegraphics[width=0.45\textwidth]{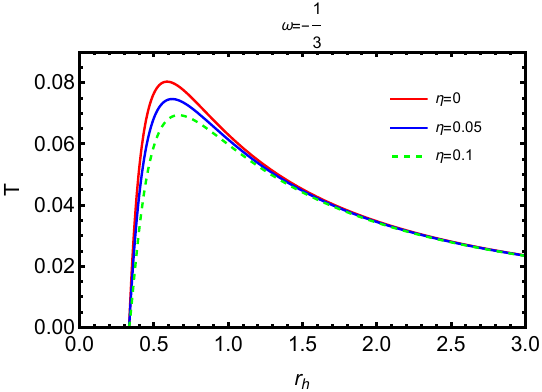}%
        \label{fig:graph1}
    }
    \hfill
    \subfloat[]{%
        \includegraphics[width=0.45\textwidth]{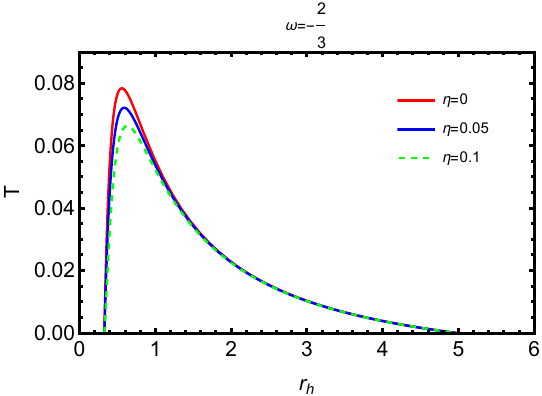}%
        \label{fig:graph2}
    }
    \\
    \subfloat[]{%
        \includegraphics[width=0.45\textwidth]{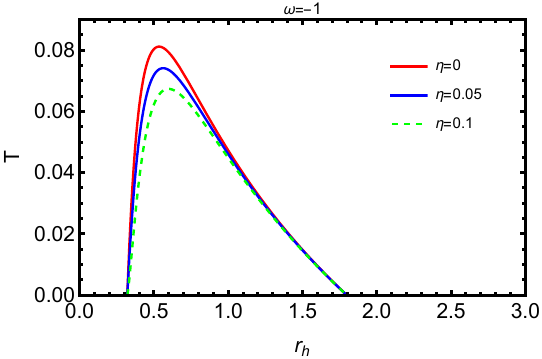}%
        \label{fig:graph3}
    }
    \caption{Temperature vs radius of event horizon $ r_{h} $ graph for $ a=0.1 $ , $ \alpha=0.1 $, $ Q=0.3 $ and $ E_{p}=1 $.}
    \label{fig:three_graphs}
\end{figure}

Comparing Eq. (72) and Eq. (78), we get
\begin{eqnarray}
f(E/E_{p})=1,\ \ \ g(E/E_{p})=\sqrt{1-\eta\left( \frac{E}{E_{p}}\right)^{n} }.
\end{eqnarray}
Using Eq. (79) in Eq. (77), we get 
\begin{eqnarray}
T_{RG}=\frac{r_{h}-M-\frac{r_{h}^{-3\omega}\alpha (1-3\omega)}{2}}{2\pi(r_{h}^{2}+a^{2})}\sqrt{1-\eta\left( \frac{1}{r_{h} E_{p}}\right)^{n} }.
\end{eqnarray}
Eq. (80) represents the RG corrected Hawking temperature of the KNBH surrounded by the quintessence. If $ \eta=0 $ in Eq.(80), RG effect is removed and gives the Hawking temperature of the KNBH surrounded by the quintessence. 

Considering the loop quantum gravity and $ \kappa $ Minkowski noncommutative space time, $ n=2 $, Eq. (80) becomes
\begin{eqnarray}
T_{RG}=\frac{r_{h}-M-\frac{r_{h}^{-3\omega}\alpha (1-3\omega)}{2}}{2\pi(r_{h}^{2}+a^{2})}\sqrt{1-\eta\left( \frac{1}{r_{h} E_{p}}\right)^{2} }.
\end{eqnarray}
The variation of the Hawking temperature ($ T_{RG} $) in relation to $ r_{h} $ is illustrated in Fig. (1) for $ \omega=-\frac{1}{3} $, $ \omega=-\frac{2}{3} $, and $ \omega=-1 $, while altering the RG corrected parameter $ \eta $. In this analysis, we set $ a=0.1 $, $ \alpha=0.1 $, $ Q=0.3 $, and $ E_{p}=1 $. We see that increasing the parameter $ \eta $, the Hawking temperature decreases.

The RG corrected heat capacity can be calculated as
\begin{eqnarray}
C_{RG}&=&\frac{\partial M}{\partial T_{GR}}=\frac{\partial M}{\partial r_{h}}\frac{\partial r_{h}}{\partial T_{GR}}\cr &=&2\pi r^{2}_{h}(r^{2}_{h}+a^{2})E^{2}_{p}\sqrt{1-\eta\left( \frac{1}{r_{h} E_{p}}\right)^{2}}\left[r^{3\omega}(a^{2}+Q^{2}-r^{2}_{h})-3r_{h}\alpha \omega \right]\cr&& \times \Big[r^{3\omega}\Big[ 2\eta \left(a^{2}(a^{2}+Q^{2})+2(a^{2}+Q^{2})r^{2}_{h}-r^{4}_{h} \right) \cr&&-r^{2}_{h}E^{2}_{p}\left(a^{4} +3Q^{2}r^{2}_{h}-r^{4}_{h}+a^{2}(Q^{2}+4r^{2}_{h}) \right) \Big] +3r_{h}\alpha \omega \cr&& \times \left\lbrace 2r^{4}_{h}E^{2}_{p}-a^{2}\eta -3r^{2}_{h}\eta +3(r^{2}_{h}+a^{2})(r^{2}_{h}E^{2}_{p}-\eta)\omega \right\rbrace \Big]^{-1}.
\end{eqnarray}

\begin{table}[ht]
\caption{The effect of the RG parameter $ \eta $ on the phase transition for different values of $ \omega $.\label{tab1}}
{\tabcolsep13pt\begin{tabular}{@{}c|c|cc@{}}
\hline
Values of  & Values of & $ r_{h} $ where phase   \\
$ \omega $&$ \eta$  & transition occurs \\
\hline
-1/3\hphantom{00} & \hphantom{0}0 & \hphantom{0}0.588535  \\
-1/3\hphantom{00} & \hphantom{0}0.05 & \hphantom{0}0.243894 \& 0.62338  \\
-1/3\hphantom{00} & \hphantom{0}0.1 & \hphantom{0} 0.321207 \& 0.666937 \\
-2/3\hphantom{00} & \hphantom{0}0 & \hphantom{0} 0.558045\\
-2/3\hphantom{00} & \hphantom{0}0.05 & \hphantom{0} 0.243894 \& 0.588535 \\
-2/3\hphantom{00} & \hphantom{0}0.1 & \hphantom{0} 0.320118 \& 0.62828 \\
-1\hphantom{00} & \hphantom{0}0 & \hphantom{0} 0.535178 \\
-1\hphantom{00} & \hphantom{0}0.05 & \hphantom{0} 0.243349 \& 0.562401 \\
-1\hphantom{00} & \hphantom{0}0.1 & \hphantom{0} 0.317395 \& 0.598335 \\
\hline
\end{tabular}}
\end{table}
The changes in heat capacity as a function of $ r_{h} $ are shown in Fig. (2) for $ \omega=-\frac{1}{3} $, $ \omega=-\frac{2}{3} $, and $ \omega=-1 $, while varying the RG-corrected parameter \( \eta \). Here we used $ a=0.1 $ , $ \alpha=0.1 $, $ Q=0.3 $ and $ E_{p}=1 $. The graph clearly indicates that a dual phase transition occurs under the RG influence, and the transition point increases with a rise in $ \eta $, whereas only a single phase transition is observed in the absence of RG effects as shown in Table 1. When the heat capacity is zero, the black hole does not exchange any heat heat with its surroundings. According to Adler et al.\cite{107}, the vanishing of the heat capacity of the black hole implies the formation of a black hole remnant. From Eq. (82), the heat capacity is found to be zero for the following values of radius of event horizon 

\begin{figure}[htbp]
    \centering
    \subfloat[]{%
        \includegraphics[width=0.45\textwidth]{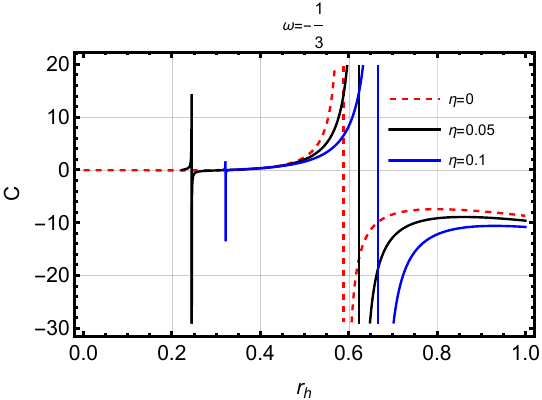}%
        \label{fig:graph1}
    }
    \hfill
    \subfloat[]{%
        \includegraphics[width=0.45\textwidth]{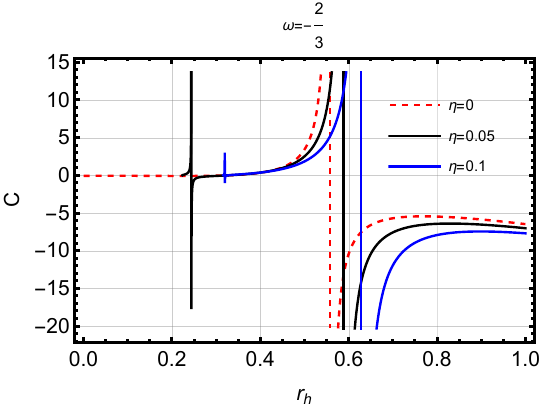}%
        \label{fig:graph2}
    }
    \\
    \subfloat[]{%
        \includegraphics[width=0.45\textwidth]{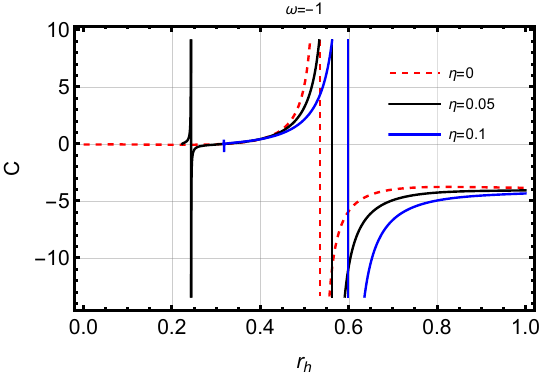}%
        \label{fig:graph3}
    }
    \caption{Heat capacity vs radius of event horizon graph for $ a=0.1 $ , $ \alpha=0.1 $, $ Q=0.3 $ and $ E_{p}=1 $.}
    \label{fig:three_graphs}
\end{figure}
\begin{eqnarray}
r_{h}=\frac{\sqrt{\eta}}{E_{p}}
\end{eqnarray}
and
\begin{eqnarray}
r_{h}&=&\sqrt{\frac{a^{2}+Q^{2}}{1-\alpha}}\ for \ \omega=-\frac{1}{3},\cr r_{h}&=&-\frac{1}{6 \Xi \alpha}\left[\Xi^{2}-\Xi +1 \right]\ for\ \omega=-\frac{2}{3}, \cr
r_{h}&=&\frac{\sqrt{\frac{1}{\alpha}\pm\frac{\sqrt{1-12a^{2}\alpha -12 Q^{2}\alpha}}{\alpha}}}{\sqrt{6}}\ for\ \omega=-1,
\end{eqnarray}
where $ \Xi =\left[ 54(a^{2}+Q^{2})\alpha^{2}-1+\frac{1}{2}\sqrt{4\left( 1-54(a^{2}+Q^{2})\alpha^{2}\right)^{2}-4 }\right]^{\frac{1}{3}}  $.
The remnant horizon radius is found to be dependent on the RG parameter ($ \eta $) and the spin($ a $), charge ($ Q $), and the quintessence parameters ($ \alpha,\omega $). Accordingly, we determine the remnant masses as 
\begin{eqnarray}
M_{rem(RG)}=\frac{1}{2}\left[\frac{a^{2}E^{2}_{p}+Q^{2}E^{2}_{p}+\eta}{E_{p}\sqrt{\eta}}-\alpha \left(\frac{\sqrt{\eta}}{E_{p}} \right)^{-3\omega}  \right] \ for \ r_{h}=\frac{\sqrt{\eta}}{E_{p}}.
\end{eqnarray}

\begin{eqnarray}
M_{rem(-\frac{1}{3})}&=&\sqrt{(a^{2}+Q^{2})(1-\alpha)}\ for \ \omega=-\frac{1}{3},\cr
M_{rem(-\frac{2}{3})}&=&-\frac{1}{72\alpha \Xi^{2}(1-\Xi +\Xi^{2})}\Big[ 1+3\Xi -6\Xi^{2}+(11 +216 a^{2}\alpha^{2}\cr&& +216Q^{2}\alpha^{2})\Xi^{3}-6\Xi_{4}+3\Xi^{5}+\Xi^{6} \Big]\ for \ \omega=-\frac{2}{3},\cr
M_{rem(-1)}&=&\frac{1}{12\sqrt{6}\sqrt{\frac{1}{\alpha}\pm\frac{\sqrt{1-12a^{2}\alpha -12 Q^{2}\alpha}}{\alpha}}}\Big[36(a^{2}+Q^{2})\cr &+& 6\left( \frac{1}{\alpha}\pm\frac{\sqrt{1-12a^{2}\alpha -12 Q^{2}\alpha}}{\alpha}\right)\cr &-&\alpha \left(\frac{1}{\alpha}\pm\frac{\sqrt{1-12a^{2}\alpha -12 Q^{2}\alpha}}{\alpha} \right)^{2} \Big]\ for \ \omega=-1.
\end{eqnarray}
Considering the quintessence, there exists a pressure described by\cite{82}
\begin{eqnarray}
P=-\frac{3}{2}\frac{\alpha \omega^{2}}{r_{h}^{3(\omega +1)}}.
\end{eqnarray}
\begin{figure}[htbp]
    \centering
    \subfloat[]{%
        \includegraphics[width=0.45\textwidth]{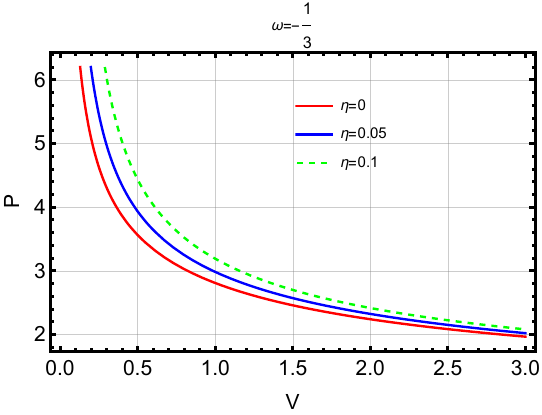}%
        \label{fig:graph1}
    }
    \hfill
    \subfloat[]{%
        \includegraphics[width=0.45\textwidth]{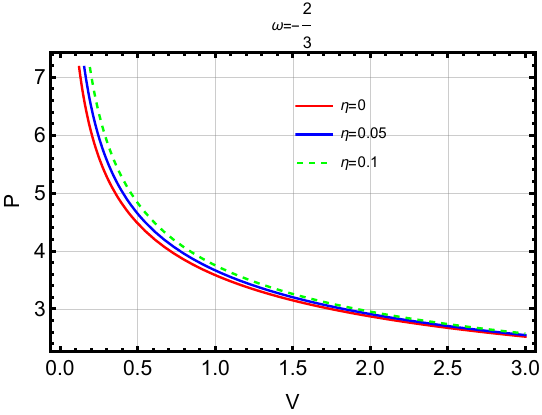}%
        \label{fig:graph2}
    }
    \\
    \subfloat[]{%
        \includegraphics[width=0.45\textwidth]{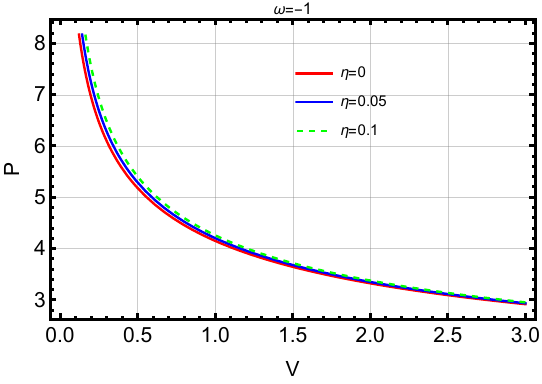}%
        \label{fig:graph3}
    }
    \caption{Pressure vs volume graph for $ a=0.1 $, $ \alpha=0.1 $, $ Q=0.3 $, $ T_{RG}=1 $ and $ E_{p}=1 $. }
    \label{fig:three_graphs}
\end{figure}
The thermodynamic volume is defined as\cite{112}
\begin{eqnarray}
V=\frac{\partial M}{\partial P}=\frac{r^{3}_{h}}{3\omega^{2}}.
\end{eqnarray}
Substituting $ r_{h}=\left(3 V \omega^{2} \right)^{\frac{1}{3}}  $ from Eq. (88) and solving $ \alpha $ from Eq. (81) into Eq. (87), we get the RG corrected pressure as
\begin{eqnarray}
P_{RG}&=&\frac{\omega}{6\times 3^{\frac{5}{6}}(V\omega^{2})^{\frac{4}{3}}}\Big[3\times 3^{\frac{1}{6}}(V\omega^{2})^{\frac{2}{3}}-\sqrt{3}(a^{2}+Q^{2})\cr&& - \frac{12\pi T_{RG} \left(3V \omega^{2}+3^{\frac{1}{3}}a^{2}(V\omega^{2})^{\frac{1}{3}} \right) }{\sqrt{3-\frac{3^{\frac{1}{3}}\eta}{E^{2}_{p}(V\omega^{2})^{\frac{2}{3}}}}} \Big].
\end{eqnarray} 
Pressure vs volume isothermal graph are shown in Fig. (3) for $ \omega=-\frac{1}{3} $, $ \omega=-\frac{2}{3} $, and $ \omega=-1 $, while varying the RG-corrected parameter \( \eta \). Here we used $ a=0.1 $ , $ \alpha=0.1 $, $ Q=0.3 $, $ T_{RG}=1 $ and $ E_{p}=1 $.

Area of the event horizon of the black hole is define as 
\begin{eqnarray}
A=\int_{0}^{2\pi}\int_{0}^{\pi}\sqrt{g_{22}g_{33}}d\theta d\phi.
\end{eqnarray} 

Under gravity rainbow, the area of the event horizon is modified as 
\begin{eqnarray}
\tilde{A}=\int_{0}^{2\pi}\int_{0}^{\pi}\sqrt{\tilde{g_{22}} \tilde{g_{33}}}d\theta d\phi &=\frac{A}{ g(E/E_{p})}.
\end{eqnarray}

Then, the RG corrected entropy of the Kerr-Newman black hole is given by
\begin{eqnarray}
S_{RG}=\frac{\tilde{A}}{4}=\frac{S_{bh}}{ g(E/E_{p})},
\end{eqnarray}
where $ S_{bh}=\frac{A}{4} $ is the original entropy of the KNBH. 

Substituting Eq. (79) into Eq. (92) and considering loop quantum gravity, we get  
\begin{eqnarray}
S_{RG}=\frac{S_{bh}}{\sqrt{1-\eta\left( \frac{1}{r_{h} E_{p}}\right)^{2}}}.
\end{eqnarray}
It is noted from the above equation that RG corrected entropy exists for $ r_{h}> \frac{\sqrt{\eta}}{E_{p}} $. If $ \eta=0 $, Eq. (93) reduces to the original entropy of the KNBH.
\section{Discussion and Conclusion}
In this paper, using the Generalized Klein-Gordon equation and the Generalized Dirac equation in the presence of quantum gravity effects under the influence of GUP, we discuss the tunneling of the scalar and fermion particles across the event horizon of KNBH surrounded by the quintessence. Using the WKB approximation the tunneling probabilities of the particles across the event horizon of the black hole are calculated. The corrected Hawking temperatures $ T_{BH} $ and $ T_{QF} $ of the tunneling of scalar and fermion particles of the KNBH is found to be dependent not only on the properties of the black hole but also on the quantum numbers of the emitted particles and quintessence parameters. $ T_{BH} $ increases or decreases according to $ \Pi<0 $ or $ \Pi>0 $ as given in Eq. (25). Likewise, according to  Eq. (46) the temperature $ T_{QF} $ increases or decreases based on whether $ \Upsilon<0 $ or $ \Upsilon>0 $. Unlike earlier models that were discussed without an exotic field, this study simultaneously incorporates rotation, electric charge, and quintessence, a dark energy model, leading to more astrophysically realistic and parameter-rich corrections.

The Hamilton-Jacobi method is based on the semiclassical WKB approximation and the separability of the action, and it has been successfully applied across a wide range of stationary spacetimes, including those involving rotating and charged black holes. While the Rarita-Schwinger approach is better suited to examining the tunneling behaviour of higher-spin particles. Tunneling of fermions is studied using the modified Hamilton-Jacobi equation derived from the modified Rarita–Schwinger equation. The dynamical equation of spin$ -\frac{3}{2} $ is governed by the Rarita–Schwinger equation, the integral half spin fermion satisfies the modified Hamilton-Jacobi equation given in Eq. (48) as there is no term affected by the specific spin. The modified Hawking temperature ($ T_{RH} $) is found to be increased from the original due to the correction term with $ \sigma > 0$ . Further, we have calculated the Hawking temperature under quantum gravity effects. Based on Eq. (64), it can be inferred that the Hawking temperature decreases due to quantum corrections, beyond semi-classical theory. From the corrected entropy of the black hole in Eq. (71), which we have calculated beyond the semi-classical theory using the modified Hamilton-Jacobi equation, it is observed that the logarithmic correction terms are linearly dependent.

We study the thermodynamics of the KNBH surrounded by the quintessence under gravity rainbow. Using the rainbow functions that are related to loop quantum gravity, we calculate the modified hawking temperature. The modified Hawking temperature ($ T_{RG} $) is observed to be influenced by both quintessence and the gravity rainbow. In the Temperature versus radius graph shown in Fig. (1), it is noted that $ T_{RG} $ decreases as the gravity rainbow parameter $ \eta $ increases, and the range where $ T_{RG} $ remains positive diminishes as the quintessence field parameter $ \omega $ decreases. Subsequently, the RG corrected heat capacity is calculated, and the possible formation of remnant mass is discussed, showing that the black hole is not completely evaporated in the presence of gravity rainbow. From the heat capacity versus radius graph depicted in Fig. 2, two phase transitions are observed, with the transition point rising as $ \eta $ increases, whereas only a single phase transition is observed in the absence of RG effects. The equation of state is also determined under the effect of RG surrounded by the quintessence. From the RG corrected P-V isotherm graph shown in Fig. (3), we can conclude that the pressure increases as $ \eta $ increases but the impact of RG appears to diminish as $ \omega $ decreases. Then, we calculate the RG corrected entropy given in Eq. (93). It is found that RG corrected entropy exists for $ r_{h}> \frac{\sqrt{\eta}}{E_{p}} $. When $ \eta=0 $, it reduces to the original entropy.

\section*{Acknowledgments}
The first author is being supported by the INSPIRE Fellowships of the Department of Science and Technology(DST),-New Delhi, India (INSPIRE Code: IF200576).

\section*{ORCID}
\noindent Aheibam Boycha Meitei - \url{https://orcid.org/0009-0001-7566-5061}
\noindent Irom Ablu Meitei - \url{https://orcid.org/0000-0001-7420-7774}

\noindent T. Ibungochouba Singh - \url{https://orcid.org/0000-0002-2568-0343}

\noindent Kangujam Yugindro Singhi - \url{https://orcid.org/0000-0002-8976-8133}

\appendix

\end{document}